\documentclass[sigconf]{acmart}

\usepackage{csquotes}
\usepackage{framed}
\definecolor{shadecolor}{gray}{0.85}
\usepackage{hyperref}
\AtBeginDocument{%
  \providecommand\BibTeX{{%
    \normalfont B\kern-0.5em{\scshape i\kern-0.25em b}\kern-0.8em\TeX}}}
\copyrightyear{2023}
\acmYear{2023}
\setcopyright{rightsretained} 
\acmConference[EASE '23]{Proceedings of the International Conference on Evaluation and Assessment in Software Engineering}{June 14--16, 2023}{Oulu, Finland}
\acmBooktitle{Proceedings of the International Conference on Evaluation and Assessment in Software Engineering (EASE '23), June 14--16, 2023, Oulu, Finland}
\acmDOI{10.1145/3593434.3593453}
\acmISBN{979-8-4007-0044-6/23/06}

\begin{document}

%%
%% The "title" command has an optional parameter,
%% allowing the author to define a "short title" to be used in page headers.
\title{Implementing AI Ethics: Making Sense of the Ethical Requirements}

\author{Mamia Agbese}
\affiliation{%
\department{Faculty of Information Technology}
 \institution{University of Jyv\"askyl\"a}
 \streetaddress{Seminaarinkatu 15, 40014}
 \city{Jyv\"askyl\"a}
 \country{Finland}}
 \email{mamia.o.agbese@jyu.fi}

\author{Rahul Mohanani}
\affiliation{%
\department{Faculty of Information Technology}
 \institution{University of Jyv\"askyl\"a}
 \streetaddress{Seminaarinkatu 15, 40014}
 \city{Jyv\"askyl\"a}
 \country{Finland}}
 \email{rahul.p.mohanani@jyu.fi}

\author{Arif Ali Khan}
\affiliation{%
\department{M3S Empirical Software Engineering Research Unit}
  \institution{University of Oulu}
  %\streetaddress{8600 Datapoint Drive}
  \city{Oulu}
  \country{Finland}}
\email{arif.khan@oulu.fi}

\author{Pekka Abrahamsson}
\affiliation{%
\department{Faculty of Information Technology and Communications Sciences}
  \institution{University of Tampere}
  %\streetaddress{1 Th{\o}rv{\"a}ld Circle}
  \city{Tampere}
  \country{Finland}}
\email{pekka.abrahamsson@tuni.fi}

\begin{abstract}
Society's increasing dependence on Artificial Intelligence (AI) and AI-enabled systems require a more practical approach from software engineering (SE) executives in middle and higher-level management to improve their involvement in implementing AI ethics by making ethical requirements part of their management practices. However, research indicates that most work on implementing ethical requirements in SE management primarily focuses on technical development, with scarce findings for middle and higher-level management. We investigate this by interviewing ten Finnish SE executives in middle and higher-level management to examine how they consider and implement ethical requirements. We use ethical requirements from the European Union (EU) Trustworthy Ethics guidelines for Trustworthy AI as our reference for ethical requirements and an Agile portfolio management framework to analyze implementation. Our findings reveal a general consideration of privacy and data governance ethical requirements as legal requirements with no other consideration for ethical requirements identified. The findings also show practicable consideration of ethical requirements as technical robustness and safety for implementation as risk requirements and societal and environmental well-being for implementation as sustainability requirements. We examine a practical approach to implementing ethical requirements using the ethical risk requirements stack employing the Agile portfolio management framework.
\end{abstract}

%% The code below is generated by the tool at http://dl.acm.org/ccs.cfm.
%% Please copy and paste the code instead of the example below.
%%
%\begin{CCSXML}
%<ccs2012>
 %<concept>
 % <concept_id>10011007.10010940</concept_id>
  %<concept_desc>Computer systems organization~Embedded systems</concept_desc>
  %<concept_significance>500</concept_significance>
 %</concept>
 %<concept>
  %<concept_id>10010520.10010575.10010755</concept_id>
  %<concept_desc>Computer systems organization~Redundancy</concept_desc>
  %<concept_significance>300</concept_significance>
 %</concept>
 %<concept>
  %<concept_id>10010520.10010553.10010554</concept_id>
  %<concept_desc>Computer systems organization~Robotics</concept_desc>
  %<concept_significance>100</concept_significance>
 %</concept>
 %<concept>
  %<concept_id>10003033.10003083.10003095</concept_id>
  %<concept_desc>Networks~Network reliability</concept_desc>
  %<concept_significance>100</concept_significance>
 %</concept>
%</ccs2012>
%\end{CCSXML}



%\ccsdesc[500]{Computer systems organization~Embedded systems}
%\ccsdesc[300]{Computer systems organization~Redundancy}
%\ccsdesc{Computer systems organization~Robotics}
%\ccsdesc[100]{Networks~Network reliability}

%%


%\received{20 February 2007}
%\received[revised]{12 March 2009}
%\received[accepted]{5 June 2009}

%%
%% This command processes the author and affiliation and title
%% information and builds the first part of the formatted document.
\maketitle

\section{Introduction}
The progress of implementing Artificial Intelligence (AI) ethics as ethical requirements in software engineering (SE) continues to witness growth, particularly as society becomes more dependent on AI and AI-enabled systems \cite{hleg2019definition}. Ethical requirements or ethical requirements for AI are requirements for AI and AI-enabled systems derived from AI guidelines, principles, or ethical codes (norms) like legal requirements derived from laws and regulations to disseminate AI ethics practices \cite{guizzardi2020ethical}. Although researchers, practitioners, regulatory bodies, and other government agencies have made significant progress at the technical developmental levels \cite{papagiannidis2021deploying,berente2021managing}, a gap still exists from portfolio management and higher management levels, likened to the middle and higher-level SE management \cite{vahaniitty2012towards,rautiainen2011towards,berente2021managing}.

Research indicates that ethical requirements are hardly considered a priority at the middle and higher SE management levels due to the low perceived effects on human lives and non-translation to financial value \cite{brendel2021ethical}. A study, \cite{morley2021operationalising}, suggested that SE executives rarely engage in pro-ethical designs, consequently confining most responsibilities to the designers and developers at teams or individual project levels. Also, product teams rarely refer to ethical requirements with executive management as they are not influential in product strategy except where legal requirements such as privacy laws are involved \cite{lynden_2020}. In such scenarios, implementation of ethical requirements at these levels of management risk being superficial — staying at a level of communications, corporate social responsibility projects, and occasional emergency interventions \cite{lynden_2020}. However, standards such as the IEEE Std 7000™-2021 advocate an "all-hands-on-deck'' approach in implementing ethical requirements towards developing trustworthy or responsible AI for society \cite{9536679, berente2021managing}. The standard proposes practical engagement of all actors, from the higher management, middle management, lower management, and all actors on AI and AI-enabled products, for implementing ethical requirements within SE businesses \cite{9536679}. The aim is to determine, address and sustain the ethical requirements of the AI product to help ascertain its significant value, i.e., its ethical requirement value (ERV),  for the stakeholders \cite{9536679}. Establishing the importance of ethical requirements may transcend finances to deepen engagement with human rights and other social values to achieve beyond legal compliance, necessitating that SE executives at these levels consider practically implementing ethical requirements as part of their management practices \cite{9536679}.

However, the discourse on ethical requirements, ERV, and practical implementation by middle and higher-level SE management are virtually non-existent. Most research focuses on determining and implementing ethical requirements at the technical development stage as functional and non-functional requirements at team levels or individual projects \cite{guizzardi2020ethical,kemell2022utilizing,halme2021write}. This gap motivates our study as we investigate how ethical requirements and their value (ERV) are considered and practically implemented in management matters by middle and higher-level SE management \cite{9536679,berente2021managing,baker2021management}. We use ethical requirements from the  European Union (EU) ethics guidelines for trustworthy AI and agile portfolio management framework by \cite{vahaniitty2012towards, rautiainen2011towards} for our investigation. Our main research question is \textit{ How do middle and higher-level SE management consider ethical requirements?} We break the main research question into five sub-questions, as follows--- \textbf{RQ1}: \textit{ What do middle and higher-level SE management understand by ethical requirements?}; \textbf{RQ2}: \textit{What value do ethical requirements represent at middle-higher level management?}; \textbf{RQ3}: \textit{How can the value of ethical requirements  be improved?}; \textbf{RQ4}: \textit{How are ethical requirements currently implemented at middle-higher level SE management?} and \textbf{RQ5}: \textit{How can ethical requirements be practically implemented?}. The aim is that answering these questions may provide initial steps to understanding how ethical requirements implementation can be further improved and contribute this knowledge to the discourse on AI ethics implementation and SE management practices. 

The following section discusses the related work for our study. Section 3 describes the research framework, methodology, data collection, analysis, and findings. Section 4 reflects on the findings, and Section 5 concludes our study.

\section{Related work} 
AI ethics generally deals with the issues raised in developing, deploying, and using AI systems and involves the moral behavior of humans in the design and development of AI \cite{muller2020ethics}. It aims to ensure that people, processes, and organizations that engineer AI assume responsibility and are engineered ethically in line with moral codes and principles \cite{muller2020ethics}. AI ethics also includes concerns about the behavior of machines and the possibility of a technological singularity from super-intelligent AI \cite{muller2020ethics}, which is outside the scope of this study. AI systems are software systems with one or more functionalities, such as speech and image recognition enabled by AI components \cite{martinez2022software}. They act in physical or digital dimensions by perceiving their environment, acquiring data, interpreting the collected data, and processing the information to decide the best course of action to achieve complex goals \cite{hleg2019definition}. However, their use of data for training and learning and their interrelation with humans introduces a new paradigm that may demand a more dedicated approach to their management \cite{martinez2022software}.

AI ethics issues such as regulation, privacy, bias, transparency, relevance, and governance have become more mainstream as technology evolves. Intervention from various governmental and private bodies culminating in AI ethics principles and policies to serve as guidelines is a step in the right direction \cite{jobin2019global,guizzardi2020ethical}. A study by \cite{jobin2019global} revealed over 80 AI ethics principles from the national and international scene where eleven overarching ethical principles emerged: transparency, justice and fairness, non-maleficence, responsibility, privacy, beneficence, freedom and autonomy, trust, dignity, sustainability, and solidarity \cite{jobin2019global}. The five principles of transparency, justice and fairness, non-maleficence, responsibility, and privacy are increasingly becoming more prominent in AI ethics issues \cite{jobin2019global}. However, AI ethics have been criticized for their ethics washing \cite{bietti2021ethics} and lack of actionable tools and practices in transitioning the inherent principles as ethical requirements \cite{vakkuri2020just,jobin2019global}, requiring a more hands-on-approach in their implementation. 

\subsection{Ethical Requirements for AI}
Ethical requirements for AI  are requirements for AI systems derived from AI guidelines, principles, or ethical codes (norms) \cite{guizzardi2020ethical}. We use the ethical requirements of the Ethics Guidelines for Trustworthy AI for our study. The guidelines use three main components of lawful, ethical, and robustness as a bedrock for trustworthy AI \cite{hleg2019definition}. The legal aspect is not addressed as the guidelines recommend that AI design, development, and management conform to the laws of the land where it is developed \cite{ai2019high}. Robustness makes AI and AI-enabled systems gain technical robustness in the society \cite{ai2019high}. 

Our study focuses on the ethical requirements derived from ethics principles. Four of them in the Ethics Guidelines for Trustworthy AI include: Respect for human autonomy, prevention of harm, fairness, and explicability, which generates seven ethical requirements \cite{ai2019high}. The requirements are: \textit{Human agency and oversight}: Focuses on fundamental human rights and the need for human agency and oversight \cite{hleg2019definition}. \textit{Technical robustness and safety}: Focuses on the robustness of AI systems to enable them to be resilient to attacks, proffer safety, and security, provide general safety and a fall-back plan and be accurate, reliable and reproducible \cite{hleg2019definition}. \textit{Privacy and Data Governance}: deals with privacy and data issues such as respect for privacy, quality of data, integrity, and access to data \cite{hleg2019definition}. \textit{Transparency}: deals with issues of traceability, explainability and communication \cite{hleg2019definition}. \textit{Diversity, non-discrimination, and fairness}: deals with unfair bias, accessibility, universal design and stakeholder participation \cite{hleg2019definition}. \textit{Societal and environmental well-being}: deals with issues such as sustainability and environmental friendliness, social impact, society, and democracy \cite{hleg2019definition}. \textit{Accountability}: deals with audit issues, reporting, and minimization of negative impacts alongside trade-offs and redress matters \cite{hleg2019definition}. A breakdown of the ethical requirements from the EU Trustworthy guideline and the underpinning AI ethics principles are presented in Table \ref{tab:freq1} .

\begin{table*}
  \caption{Ethical Requirements}
  \label{tab:freq1}
  \begin{tabular}{ccccl}
    \toprule
    \# & AI ethics principles \cite{jobin2019global} & Ethical requirements (EU Trustworthy guideline) \cite{hleg2019definition}&Underpinning ethical principles \cite{jobin2019global}\\
    \midrule
     \ 1&Transparency &Human agency and oversight& Freedom and autonomy, Dignity\\
    \ 2 &Justice and fairness& Technical robustness and safety & Non-maleficence, Dignity\\
    \ 3 & Non-maleficence&Privacy and data governance & Privacy\\
    \ 4 &Responsibility &Transparency & Transparency\\
    \ 5 &Privacy& Diversity, Non-discrimination, and fairness & Justice and fairness\\
    \ 6 &Beneficence& Societal and environmental well-being & Beneficence\\
    \ 7 & Freedom and autonomy&Accountability& Responsibility and accountability\\
    \ 8& Trust& & \\
    \ 9 & Sustainability& &\\
    \ 10 &Dignity&&\\
    \ 11& Solidarity& &\\
  \bottomrule
\end{tabular}
\end{table*}
 
\subsection{Implementing  Ethical Requirements for AI in Software Engineering Management}
SE management involves applying management activities and practices to ensure that products and services are delivered efficiently and effectively to the benefit of stakeholders \cite{freeman2001software}. At technical development management levels in SE, ethical requirements are considered functional and non-functional ethical requirements \cite{vakkuri2020current,guizzardi2020ethical}, however within middle and higher level management, virtually no representation of ethical requirements seems to exist \cite{mokander2021ethics}. Martinez-Fernandez et al. \cite{martinez2022software} highlight this limited synthesized research on SE approaches to designing, developing, and maintaining or managing AI systems in their analysis on state of the art on SE knowledge on SE and AI systems. They classify their findings according to Software Engineering Body of Knowledge (SWEBOK) \cite{martinez2022software}. Their results reveal a disparity, with studies focused on the dominant technical aspects and the maintenance and management aspects appearing neglected \cite{martinez2022software}. This direction could be a consequence of the speed of the empirical multi-disciplinary bottom-up approach of research, which has primarily targeted the technical needs and challenges of AI developers and stakeholders with a scarce contribution to the managerial side \cite{baker2021management,vulpen2018continuous}.
 
 But Berente et al. \cite{berente2021managing} explain that due to the speed of investment in AI worldwide, SE management does not have the luxury of catching up on ethical requirements-related issues as they run the risk of these issues spiraling out of their control if not addressed effectively and timely. They emphasize the imperativeness for middle and higher-level SE management to proactively lead the charge on implementing ethical requirements to aid the value sustenance of their AI \cite{berente2021managing}. Also, The IEEE Std 7000™-2021 \cite{9536679} explains that socially responsible SE organizations understand the significance of their decisions and actions on society and not just on their financial bottom line. The standard explains that direct organizational stakeholders, from the top management down to the team levels, should be invested in implementing ethical requirements \cite{9536679}. While implementing ethical requirements is not the sole responsibility of top management, they play a significant role in setting expectations for their implementation and verifying control of performance and outcomes \cite{9536679}. The IEEE Std 7000™-2021 explains that elicitation of ethical requirements can be carried out using ethical considerations such as guidelines, frameworks, or practices from top management down to team levels \cite{9536679}. Ethical requirements are described in terms of the value they provide the organization, and value as the significance that influences the decision or selection of AI projects. Using ethics considerations for elicitating value is encouraged as it can result in a case for ethics to be analyzed across the different groups within organizations and help interpret the value attached to the AI project. The outcome value can then be further investigated and prioritized with the concurrence of management and cascaded down to the activities and tasks to help develop the AI system \cite{9536679}.
 
\section{Research Methodology}
We follow an exploratory approach to help determine the study's primary aim. The exploratory process provides flexibility to enable adaptability as the investigation unfolds, mainly where little information in the research exists \cite{swedberg2020exploratory}.
\subsection{Data Collection}
We used semi-structured interviews and open-ended questions to collect our data by interviewing ten SE executives from ten different Finnish software companies in Table \ref{tab:freq}. Due to the challenge of gaining access to this group of interviewees' network, we used a snowballing approach. Two of the authors with industry contacts enabled the first set of interviews. After gaining access to the first set of interviewees, they introduced us to subsequent interviewees or contacts in the same field. This way, we were able to contact 13 interviewees in total. However, only ten responded, with three unable to participate. In this way, we focused on personnel directly related to the research \cite{parker2019snowball,harrell2009data}. The interviewees had varying degrees of experience in SE management with direct and indirect management of AI and AI-enabled solutions and systems. We conducted nine interviews over Zoom and one face-to-face. Each session lasted 30-60 minutes and was recorded by the interviewer for transcription. We framed the interview questions by consulting frameworks of \cite{jugend2014product,cooper1999new}.

\begin{table*}[h]
  \caption{Interviewee context}
  \label{tab:freq}
  \begin{tabular}{cccl}
    \toprule
    Interviewee number & Background & Position\\
    \midrule
    \ Interviewee 1&Software consumer electronics&Executive\\
    \ Interviewee 2& ICT solutions&Consultant\\
    \ Interviewee 3&Data management&Consultant\\
    \ Interviewee 4&IT services&CEO\\
    \ Interviewee 5&Software services&CEO\\
    \ Interviewee 6&Video surveillance systems&Management\\
    \ Interviewee 7&Digitization and ICT&Product Management Director\\
    \ Interviewee 8&Digitalization and ICT &Management\\
    \ Interviewee 9&Software quality and measurement&Consultant\\
    \ Interviewee 10&Software IT and telecommunications&Management\\
 \bottomrule
\end{tabular}
\end{table*}

\subsection{Interview Protocol}
The interview used an inverted funnel approach which typically begins with background questions and builds up to more open-ended questions enabling respondents to be more comfortable before being asked broader questions  \cite{harrell2009data}. The sessions started with the interviewer's introduction, the purpose of the research, and the reason for the respondent's participation. The interviewer then explained the semi-structured nature of the interview, estimated length of time, recording of the interview, intended usage of the data, anonymization of the data, the chance of the interviewee accessing the data, and options not to participate or have data removed if they chose. One of the interviewees, a practitioner from the authors' research group, served as a test for the interview to clarify unclear terminologies or questions that may be inappropriate for the audience\cite{harrell2009data}. The test was included in the data set as virtually no improvements were made to the interview questions \cite{harrell2009data}. The interview questions can be found here: \href{https://doi.org/10.5281/zenodo.7827529}{Interview questions}. 
\subsection{Analysis}
The interviews were transcribed manually, and using thematic analysis, we identified, analyzed, and reported analytic patterns or themes within the data \cite{inproceedings}. The coding process was performed manually by repeatedly reading and making notes on interview transcripts to interpret various aspects of the research topic \cite{inproceedings}. The resulting codes and the related extracts were then scrutinized and combined to repeatedly form overarching themes to ensure that the final thematic output met the research aim. The themes were built by identifying structures within the data with an explanatory capacity for the study. The first author carried out the coding. Six main themes emerged: understanding of ethical requirements, value of ethical requirements, importance, influencing factors, implementation, and visibility of ethical requirements. To avoid bias, the second researcher reviewed the codes and themes that emerged to provide an accurate representation of the data. Both authors agreed that the themes---value of ethical requirements and importance---overlapped, sharing similar content due to code saturation making it appropriate to combine them into one theme and narrow the main themes to five to ensure clarity and distinctiveness \cite{inproceedings,hennink2022sample}. The code saturation was achieved at interviewee 8.

The themes (T) are as follows: The first theme---\textit{understanding ethical requirements}, examines the understanding of ethical requirements of these management groups through the lens of Ethics Guidelines for Trustworthy AI and matches RQ1. RQ1 is motivated by the need to understand how ethical considerations are perceived and incorporated into SE management practices driven by concerns about the potential ethical implications of software development. Poor understanding of ethical requirements by these stakeholders can lead to them overlooking or downplaying their significance, leading to ethical violations or negative consequences. The second theme---\textit{value of ethical requirements} examines the value or significance (ERV) attached to ethical requirements at these levels and matches RQ2. RQ2 is motivated by the need to understand the importance and impact of ethical requirements in various SE management practices resulting from concerns.The findings could potentially inform the development of ethical initiatives for SE organizations that better reflect the values and priorities of stakeholders and help promote responsible and sustainable practices. 

\begin{figure}[htbp]
  \centering
  \includegraphics[width=\linewidth]{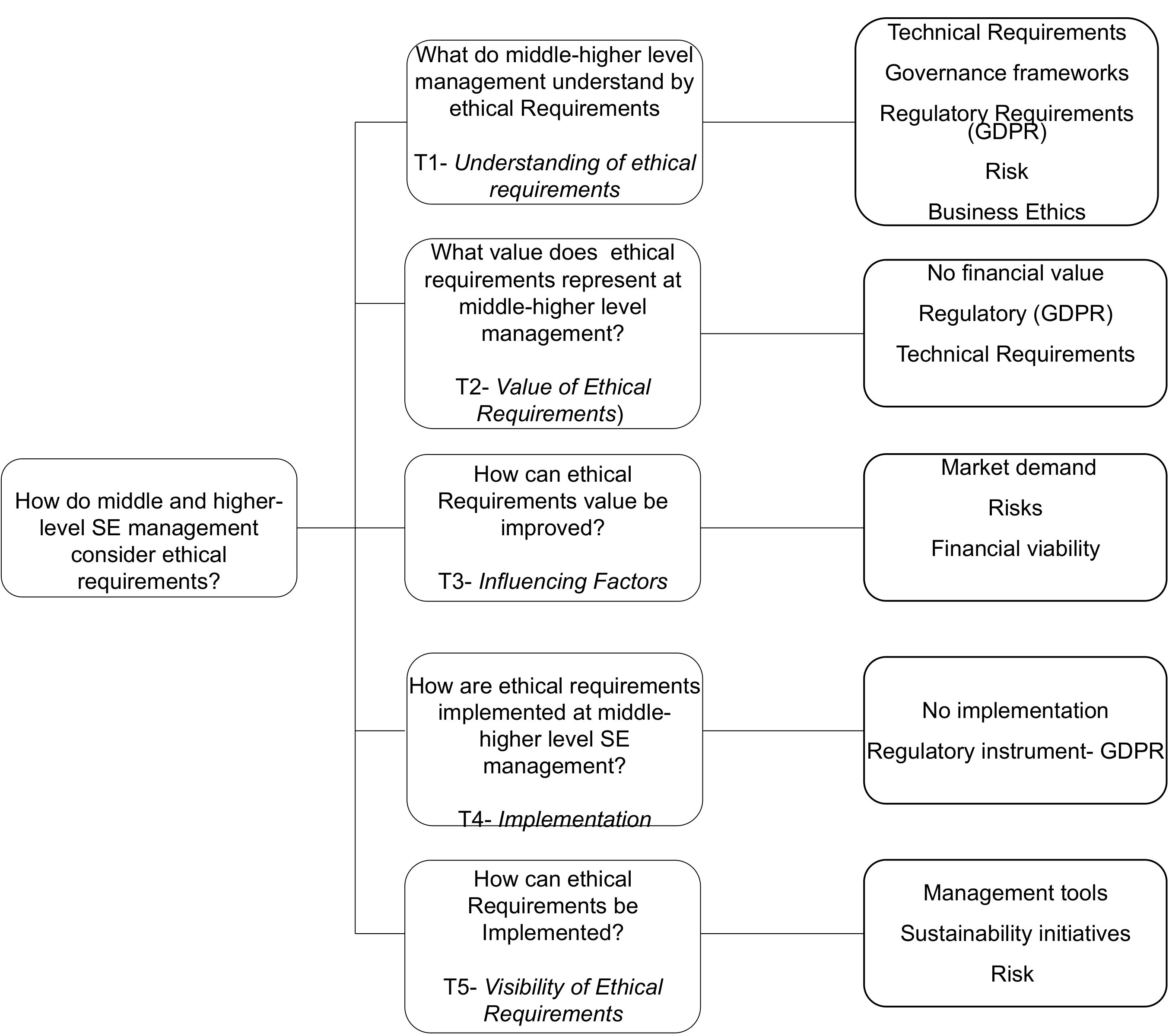}
  \caption{Thematic analysis}
  \Description{Thematic analysis}
  \label{fig:thematicanalysis}
\end{figure}

The third theme---\textit{influencing factors} examines factors that influence management practices and how the value of ethical requirements can be improved to be at par with these metrics; the theme matches RQ3. RQ3's motivation stems from the need to understand SE management practices concerning ethical requirements value better and identify improvement opportunities. Exploring this RQ can help identify areas or initiatives where ethical considerations and their value may be overlooked or inadequately addressed and provide insights on enhancing them. The fourth theme---\textit{implementation} examines the current implementation of ethical requirements, if any, and matches RQ4. RQ4 is explored to understand current implementations of ethical requirements, identify areas of improvement for ensuring that SE practices are conducted responsibly and sustainably, and for protecting the rights and interests of stakeholders. The fifth theme---\textit{visibility of ethical requirements} investigates current visibility management metrics and how ethical requirements can be implemented to match them and fits RQ5. RQ5 explores practical strategies for integrating ethical considerations into SE management practices, as it could help identify best practices for incorporating ethical considerations. The emergent themes (T) are represented as the research questions in Figure \ref{fig:thematicanalysis} and explained in the findings section.

\subsection{Findings}
The findings from the analysis are highlighted as Primary Empirical Contributions (PECs). We use quotes from the interviewees (I) to elaborate on the topic further.
\subsubsection{What do middle-higher level management understand by ethical requirements?} To answer this question, we present the interviewees' responses regarding their interpretation of ethical requirements. Similarity to business ethics was identified, suggesting that these management groups equate ethical requirements to normative business ethics or moral practices in general. Indicating that these management groups may have no particular consideration of ethical requirements \cite{berente2021managing}. 

\begin{displayquote}
    
    \textit{"I do not have any strict principle like these are the AI ethics I have, I do not claim to have extensive knowledge specific to AI, but I always try to operate within the \textbf{moral principle} that we have"}, (I4).
\end{displayquote} 
 
Another sub-theme that emerged is requirements engineering, which implies that executives view ethical requirements in a similar light to technical requirements. Extracts from interviews further elaborate on this point.
    
    \begin{displayquote}
    \textit{"I don’t see it that much different from things that have been, it’s more like how we can emphasize some areas, for example, \textbf{non-functional} things like reliability, how well we can document and explain the \textbf{functionality} of a system"}, (I1). 
    \end{displayquote}
    
    \begin{displayquote}
        \textit{"And AI and ethics especially is an \textbf{ilities}." (I10)}
    \end{displayquote}

Ethical requirements are understood as GDPR.

\begin{displayquote}
    \textit{"For example, \textbf{GDPR} is something that we strictly follow in our principles. So, we check that will the \textbf{GDPR} rules be applied, or do we explain to the clients?" } (I5)
\end{displayquote}

The risks lack of implementation can pose to the business and customers. One of the interviewees explains:

\begin{displayquote}
    \textit{"I will say that it is something everybody should consider whether you want to do that or not but especially for the decision makers that it is important that you are in ethical business, then you should consider all the \textbf{ethical risks} that are coming from your actions."} (I9).
\end{displayquote}

Ethical requirements understood in terms of governance frameworks.

\begin{displayquote}
    "\textit{There are already those \textbf{governance} structures and also much of those questions and considerations are already in place, it is just how to draw a line in between the highly valued principles that we have and already existing \textbf{governance} systems or \textbf{frameworks} or operations"} (I3)
\end{displayquote}

%\begin{tcolorbox}
\begin{shaded}
\textit{PEC1: Middle and higher-level SE management have a fragmented understanding of ethical requirements.}
\end{shaded}
%\end{tcolorbox}
 
 \subsubsection{What value does ethical requirements represent at middle-higher level management?} According to the findings, these requirements are considered relatively new and have had little impact on the market, meaning they currently have no financial value due to a lack of customer demand. One of the interviewees explains:
   
   \begin{quote}
   \textit{"It depends on \textbf{the market}. It all comes down to what the \textbf{customers want}, and if they don’t worry about these things and if \textbf{they don’t demand} you to make strong ethical consideration in your product development, if your customers \textbf{do not demand} for that then they are not ready to pay for that, and then I am not ready to build it and \textbf{demand} for that."} (I2)
   \end{quote}
   
The value of ethical requirements is assessed in terms of regulatory measures like GDPR. 
    \begin{quote}
    \textit{"We always think the \textbf{legal perspective} and the guideline is \textbf{GDPR}, we follow those rules that are in there so that’s why I think its in our everyday job."} (I6) 
     \end{quote}

This interviewee explains the value of ethical requirements as technical requirements.

\begin{displayquote}
    \textit{"First it was getting the specifications and the \textbf{requirements}, whether they were \textbf{functional or non-functional} from the business guys and the \textbf{technology} guys. And then coming up with a solution concept of how we are now dealing with these \textbf{requirements} in and how we can now turn them into \textbf{technological} or process or both of them, product and service."} (I3)
\end{displayquote}

\begin{shaded}
    \textit{PEC2}: \textit{Ethical requirements have value as technical and regulatory requirements but no financial value.}
\end{shaded}

\subsubsection{How can the value of ethical requirements  be improved?} Our analysis identified three factors that can help boost their value. One of the most significant factors is financial viability, which can be used to leverage ethical requirements and improve their monetary value. By demonstrating the economic benefits of ethical requirements, businesses can generate sales and improve their bottom line."

 \begin{quote}
   \textit{"Of course, the \textbf{profitability} of the product portfolio as a business, that is one key driver"} (I5);
   \end{quote}
   
   \begin{quote}
   \textit{"But unless they are in my two, \textbf{finance} and road map and I did not give ethical consideration because it shows that in decision-making they are not in that high level; and maybe they should be." (I1)}
   \end{quote}
   
Market demand:

\begin{quote} 
    \textit{"The only reason, two reasons why companies think about green values right now is that their \textbf{customers demand} it and a little bit of legislation like the government and the taxes and all that".} (I1)  
    \end{quote}
 
   \begin{quote}
   \textit{"Of course, in general, we made the decision to create the software was the \textbf{market potential}"} (I3)
   \end{quote}

The risks aspect: 
    
   \begin{quote}
    \textit{"but especially for the decision makers that it is important that you are in ethical business, then you should consider all the \textbf{ethical risks} that is coming from your actions."}, (I4).
     \end{quote}
     
\begin{shaded}
   \textit{PEC3}: \textit{Ethical requirements value can be enhanced as financially viable sustainability initiatives and ethical risk requirements.}
\end{shaded}

 \subsubsection{How are ethical requirements implemented at middle-higher level SE management?} According to our analysis, ethical requirements are typically implemented as legal and regulatory requirements, such as those outlined in the GDPR.
 
 \begin{displayquote}
     "\textit{Yeah we have a lot of enterprise customers and software built for enterprise class companies so the \textbf{GDPR} is almost like an everyday thing for us}". (I8)
 \end{displayquote}

 \begin{displayquote}
     "\textit{That’s kind of a question because we deal with \textbf{GDPR} almost every day, so it’s like a very normal thing for us and its like everywhere, so how could it show. Always when we make like some technology decisions, build strategies or then deep dive into certain features development, we always think the \textbf{legal} perspective and the guideline is \textbf{GDPR}, we follow those rules that are in there so that’s why I think its in our everyday job.}" (I4)
 \end{displayquote}
 
 \begin{displayquote}
    "\textit{Well, it’s a hygienic issue, its \textbf{regulatory} issue and all the big companies just have to follow it, it is the \textbf{law}. So I think that’s the reason because if you are a big company and do not follow rules, you cannot exist in this democratic world we are living in}." (I6)
\end{displayquote}
 
 However, an interviewee expressed concerns about implementation as legal requirements
 
\begin{quote}
\textit{"I don’t believe in \textbf{government intervention} with all these things. I am a little cynical but that is more like a political thing. I believe that the consumers and the customers have the ultimate power in the modern society. "} (I1)
\end{quote}
 
\begin{shaded}
     \textit{PEC4}: \textit{Ethical requirements are implemented as legal requirements.}
 \end{shaded}

\subsubsection{How can ethical requirements be practically implemented?} The analysis delves into practical ways of implementing ethical requirements beyond regulatory demands. Notably, financially viable sustainability initiatives can form the basis of sustainability requirements, while ethical risk initiatives can serve as risk requirements. These measures can be integrated into existing management tools for optimal impact.

\begin{displayquote}
         "\textit{It can be but then it has to be the hooked like checklist tools and probably to those frameworks that are dealing with the \textbf{risks}, \textbf{business risks} and \textbf{risks} involved that is compliance thing or whatever or \textbf{market risks} and those types of \textbf{frameworks} are most natural to that, I think}." (I10)
     \end{displayquote}
     
     \begin{displayquote}
         "\textit{\textbf{Risk management} is very important, I think that these principles they are nothing new, they are already there, it is about how to highlight the different aspects}." (I4) 
     \end{displayquote} 

For financial viability, ethical requirements to be marketed like sustainable initiatives is strongly identified in the analysis. The interviewees explain:

\begin{displayquote}
    "\textit{For example there is a difference between the coal energy companies and \textbf{green energy companies}. So, in a similar way, with AI ethical tools. If you are a company who use without thinking about any precautions or companies who rarely evaluate the ethical sides, that could be a nice differentiation. It adds value in the moral sense and also it can add value in the business sense if you kind of use it wisely and advertise it nicely. So it can also add business value for the company."} (I1)
\end{displayquote}

\begin{quote}
\textit{" I was talking about the \textbf{green values} and \textbf{environmental aspects} in the decision-making of companies. And unless companies take \textbf{green values} like \textbf{environmental values} as a first-class citizens, on the same level as performance, nothing is going to happen."} (I4)
\end{quote}

\begin{shaded}
   \textit{PEC5}: \textit{Ethical requirements can be implemented as ethical risk and sustainability requirements using customary frameworks.}
\end{shaded}

\section{Discussion} 
\subsection{Overview} We discuss the PECs and their implications in this section. 
\subsubsection{ PEC1 - Middle and higher-level SE management have a fragmented understanding of ethical requirements:} Ethical requirements are understood as technical or developmental requirements and the risk they can pose to SE business. These components correspond with the ethics guidelines' technical robustness and safety requirements for trustworthy AI. The executives have a divergent understanding of ethical requirements. However, they agree that their mismanagement can lead to havoc on business processes or profit loss. Despite these concerns, AI risk impact is still considered low and misunderstood. The multi-stakeholder participation and the relative newness of the technology provide no proper governance and allow for ethics washing \cite{morley2021operationalising}. This approach may explain why firms such as Volvo (defeat device story) have faced dire brand and financial consequences due to negligence in implementing ethics \cite{barn2016you}. However,  Morley et al. \cite{morley2021operationalising} advise that leaders who aim to prevent or mitigate anticipated consequences from their AI and AI-enabled systems can seek to understand the risk aspects of ethical requirements.

Ethical requirements are understood as Governance frameworks and Regulatory requirements, such as GDPR, corresponding with the Privacy and data governance requirements of the ethics guidelines for trustworthy AI. Representing ethical requirements as regulatory instruments may be effective but does not allow for a holistic representation of ethical requirements \cite{morley2021operationalising}. A study by \cite{sirur2018we} explains that one of the main reasons for companies' preparedness for GDPR is down to "fashion and explicitly enforced regulation" [p.7]. They explain that while some companies, mainly larger companies with resources, indicate an understanding and preparedness for GDPR, there is a disparity in knowledge for smaller firms. Smaller firms with less budget and understanding of GDPR struggle with compliance and only do so due to the stiff penalty of defaulting \cite{sirur2018we}. Morley et al. also \cite{morley2021operationalising} argue that this may not be the right approach as merely making AI products or services legally compliant does not necessarily make them ethically sound and socially acceptable.

Understanding ethical requirements as business ethics may stem from the Accountability component of the Ethics guidelines for trustworthy AI \cite{hleg2019definition}. The normative business ethics for similar technologies are utilized alongside management frameworks for ethical requirements \cite{baker2021management}. Baker \cite{baker2021management} explains that while managers at the executive level agree to some general understanding of ethical requirements, their knowledge is not based on AI ethics principles but more broadly on business ethics. Business ethics which most companies rely on for their ethics, is usually implemented implicitly; however, AI ethics principles go beyond normative business ethics and need an explicit representation for proper implementation \cite{baker2021management}.

\subsubsection{PEC2 - Ethical requirements have value as technical and regulatory requirements but no financial value:} These components correspond to the technical robustness and safety components, and privacy and data components in the ethics guidelines for trustworthy AI. Morley et al. \cite{morley2021initial} constructed a typology of methods and practices to assist developers in implementing AI ethics at each stage of machine learning development. While the list is extensive, the practices are somewhat consigned to the micro or development level. Similarly, \cite{ayling2021putting} reviewed the best practices and frameworks for implementing AI ethics and revealed that the tools with the most impact are those used on the production side of AI. They note that users such as developers, quality assurance personnel, and delivery roles had the most interaction and usage of AI ethics tools, with outputs passed on to executives and senior staff \cite{ayling2021putting}. However, it does not discuss the tools available for executives, lending strength to executives' reasoning at these levels of ethical requirements being technical.

Conversely, The financial or  Business value of ethical requirements is practically non-existent. A primary reason for the lack of visibility of ethical requirements could be its lack of profitability \cite{hagendorff2020ethics}. Business value does not directly correspond to ethical requirements in the Ethics guidelines for trustworthy AI; as such, PEC2 discusses components of ethical requirements like Societal and environmental well-being that can be harnessed profitably. Brendel and Vaha \cite{brendel2021ethical, vaha2011software} lend credence to this perspective as their study highlights ethical requirements' ineffectiveness at the macro level attributed to financial unprofitability. However, Hagendorff \cite{hagendorff2020ethics} explains that company managers are eager to make a profit on AI systems, with motivation based mainly on profitability logic and not on principles. He explains that while business executives may be willing to exploit AI for its economic benefits, they are unwilling to implement AI ethics principles. Also, since ethical requirements currently lack enforcement mechanisms, most businesses voluntarily ignore them in implementation \cite{hagendorff2020ethics}. As such, caution is needed in society in entrusting the implementation of AI ethics to companies \cite{hagendorff2020ethics}.

\subsubsection{PEC3 - Ethical requirements value can be enhanced as financially viable sustainability initiatives and ethical risk requirements:} The PEC analyzes how the value of ethical requirements can be improved by making it financially viable by productizing and commercializing its sustainability requirements to increase market demand. PEC3 explains that ethical requirements harnessed economically as sustainable solutions that advocate for environmentally friendly AI solutions that contribute to a fairer and equal society \cite{jobin2019global} can improve its market demand and profitability. Jacobsen et al. \cite{jacobsen2020towards} list productization as a means of achieving sustainability even in AI businesses. They explain that including sustainability practices such as productization can help businesses strengthen employer brands, scale resources, increase business opportunities, increase customer demand and help businesses maintain a reputation as innovative. \cite{jacobsen2020towards}. These components correspond with Societal and environmental well-being and technical robustness and safety requirements of the ethics guidelines' for trustworthy AI \cite{hleg2019definition}.

Similarly, SE businesses can increase the value of AI ethical requirements by positioning their products as ethical risk requirements to help mitigate AI failures and fatal outcomes that pose a business risk. Such practices may also serve as a competitive advantage to increase market value. However, Ziesche \cite{ziesche2021ai} explains that the importance of ethical requirements should align with human values and transcend financial value; as such, executives need to include these aspects in assessing ethical requirements. Hence, if the principal value of AI ethics is profit, it may indicate that the societal impact of AI ethics is far from being understood by executives.

\subsubsection{PEC4 - Ethical requirements are implemented as legal requirements:} The PEC reveals that implementation as legal requirements at middle-higher level management attributes to the high compliance and impact of legal and regulatory requirements, such as GDPR, at these levels \cite{floridi2018soft}. Floridi \cite{floridi2018soft} explains that the ethical, legal, and social implications of the GDPR implicitly incorporated in the law help to exercise discretion and adjudication on SE organizations. He explains that regulations such as GDPR provide a dual advantage. On the one hand, it determines what should and, on the other, what should not be for organizations that develop technologies such as AI, allowing them to take advantage of a calculated social advantage of the technology within predefined boundaries \cite{floridi2018soft}.

\subsubsection{PEC5 - Ethical requirements can be implemented as ethical risk and sustainability requirements using customary frameworks:} PEC5 analyzes ethical requirements implementation as sustainability initiatives and ethical risk using existing frameworks for easy adoption at these levels. The discussion regarding these initiatives is analyzed in PEC3. But regarding performance using customary practices and frameworks, Morley et al. \cite{morley2021ethics} explain that ethical requirements must be operationalized to be valid for practitioners and customers. As such, they propose implementing ethical requirements of AI at an appropriate level of abstraction where the mechanisms are not too flexible or strict, incorporating all actors in a familiar environment to enable easy adoption \cite{morley2021operationalising}. As such, we examine the use of customary practices using the ethical requirements stack.

\subsection{Ethical Requirements Stack}
We explore Agile portfolio management as a management framework in our study \cite{vahaniitty2012towards,rautiainen2011towards}. Portfolio management operates as coordinated management comprising one or more portfolios of products to achieve organizational strategies and objectives \cite{hyvari2014project}. It is also one of the essential areas of businesses that links strategy to product development \cite{vahaniitty2012towards}. From the SE angle, \cite{van2006creation} describes portfolio management as involving decision-making on existing SE products or services, the introduction of new products or services depending on market trends, product development strategy, decisions on product lifestyle, and the establishment of partnerships and contracts \cite{van2006creation}. 

\begin{figure}[h]
  \centering
  \includegraphics[width=\linewidth]{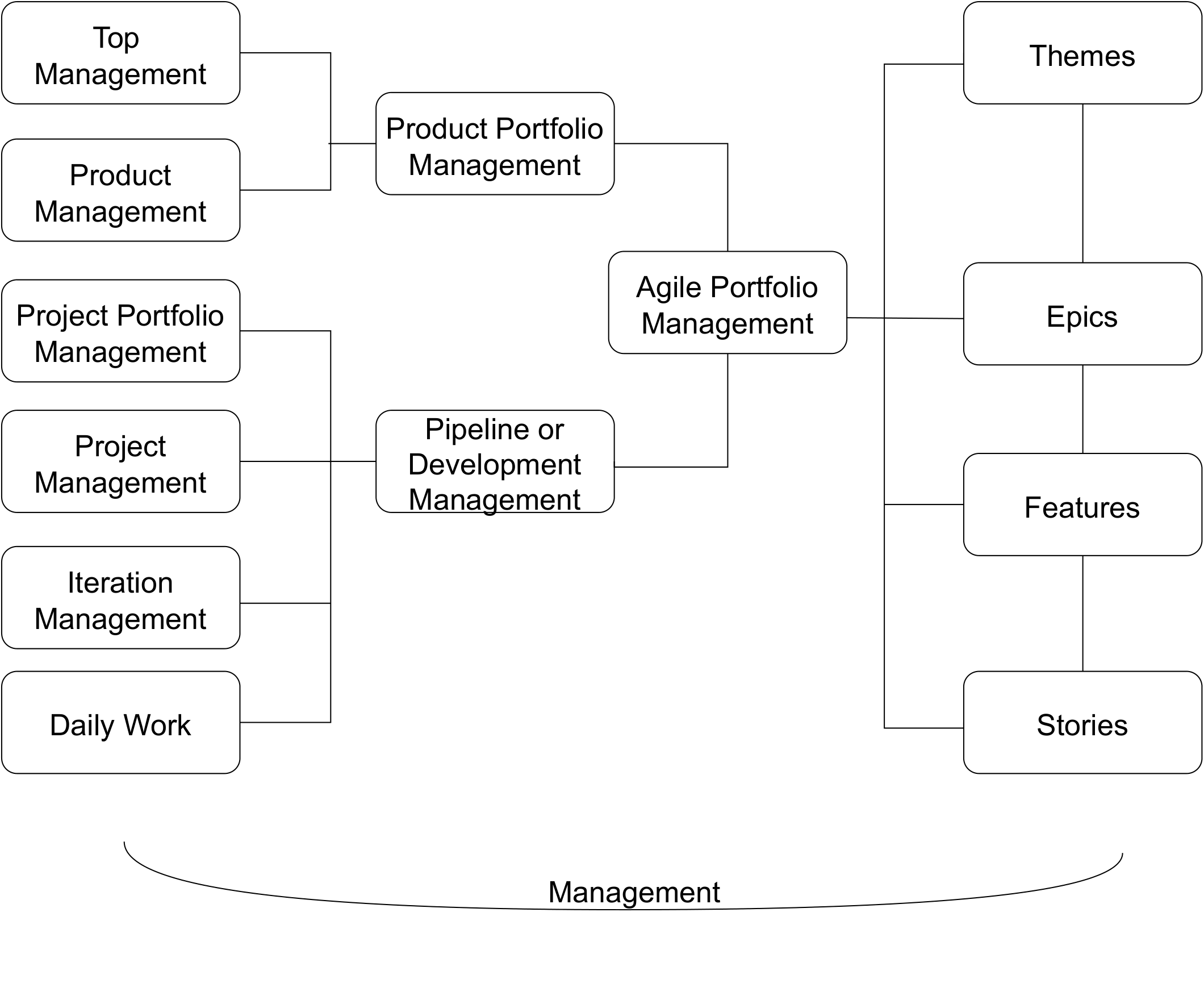}
  \caption{Agile portfolio management 
  \cite{rautiainen2011towards,vahaniitty2012towards}
    }
  \Description{Agile portfolio Management  \cite{rautiainen2011towards,vahaniitty2012towards}}
  \label{fig:Agileportfoliomanagement}
\end{figure}

 Vahaniitty et al. \cite{vahaniitty2012towards} describe Agile portfolio management using a framework based on Agile scrum practices for a practical approach to SE management. They analyze product portfolio management (portfolio management) with development portfolio management (product development or pipeline management) practices as an Agile portfolio management framework. Agile portfolio management processes spread across actor groups of top management (strategy level activities), middle management (product management activities), and operational or development or teams level \cite{rautiainen2011towards}. Agile portfolio management is described as an enterprise top-level activity product or investment theme where strategic product themes are defined and cascaded into Agile methodology concepts of epics, features and stories \cite{rautiainen2011towards}. Agile portfolio management is not just a singular process but a multi-layered approach with one layer connected to the other to achieve the organization's primary strategy. The steps are equated with corresponding agile Scrum method practices of themes, epics, features, and stories, as illustrated in Figure \ref{fig:Agileportfoliomanagement}. Agile practices are viewed as accommodating to dynamic requirements, enabling collaboration between the business and customers, and supporting early product delivery, which is vital in developing AI systems \cite{huo2004software}. Criticism of the approach includes the process being too fast-paced, which may lead to inefficient management, fragmented results, and an unclear contribution to organization goals \cite{rautiainen2011towards}. However, the changing nature of development and associated technologies and increasing communication requirements necessitates exploring such an approach to help organizations meet the market's ever-changing demands.

\subsubsection{Ethical requirement stack:} 
\begin{figure}[htbp]
  \centering
  \includegraphics[width=\linewidth]{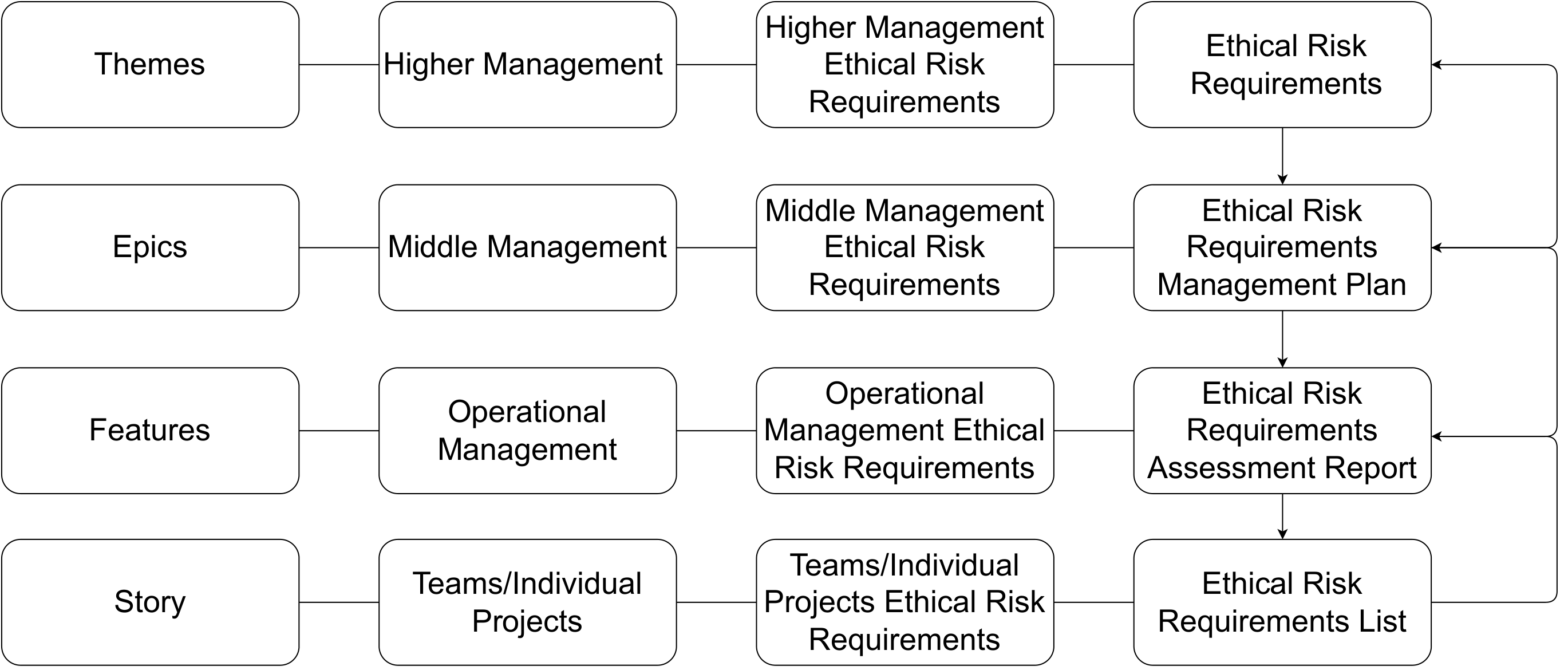}
  \caption{Ethical risk requirements  stack} 
  \Description{Ethical risk requirements  stack}
  \label{Ethicalriskrequirementsstack}
\end{figure}

We examine the use of customary practices for a practical management approach for implementing ethical requirements following the practices of the Agile portfolio management framework \cite{vahaniitty2012towards} to create an ethical requirements stack. The ethical requirements stack uses one of the recommendations for risk as ethical risk requirements, as suggested in the analysis of the study. The executives in the study seem to converge on the risk aspect, either market or enterprise risk AI can pose. Risk is akin to preventing harm (non-maleficence) of the EU AI ethics guidelines. We, therefore, examine the implementation of ethical risk requirements and explore how it can be represented at different layers of management as an ethical risk requirements stack in Figure \ref{Ethicalriskrequirementsstack}.

The executive or strategy layer creates the vision for the ethical risk requirements approach for the business. This layer corresponds with the themes in the Agile portfolio management framework \cite{vahaniitty2012towards,rautiainen2011towards}. The ethical risks are determined or outlined and communicated in terms of their percentage contribution to the organization \cite{vahaniitty2012towards,rautiainen2011towards}. The business risks at this level are defined and identified as a central theme of ethical risk requirements. The middle management layer, which corresponds to the epic layer, is responsible for cascading or interpreting the ethical risk requirements vision and managing its development in releasing the strategy of the business \cite{vahaniitty2012towards,rautiainen2011towards}. The epic layer can create an ethical risk requirements management plan in conformance with organizational policies and objectives where information and policies from top management are outlined with proper guidance for the lower layers for execution \cite{vahaniitty2012towards,rautiainen2011towards}. The epic layer involves the tactical prioritization and resource allocation across the activities competing for the same resources; it is also responsible for interpreting the themes \cite{vahaniitty2012towards}. At this level, SE businesses can engage in strategic activities to help them deliberate the ethical requirement consideration of their AI products to determine the value it presents to them \cite{9536679}. At this layer, the themes are simplified into what can translate into ethical risk requirements management plan. The ethical risk requirements management plan can be presented as a roadmap to enable the business to track progress. The resulting ethical risk requirements management plan can then be broken down into manageable practices identified in the Agile context as features cascaded to the operations levels.

The operations layer is identified as the features layer and aligns business strategy to mandatory requirements \cite{horlach2019agile, vahaniitty2012towards}. The features layer focuses on business resource allocation to deliver the bare minimum marketable features that can provide value for the business in conformity with the vision \cite{rautiainen2011towards}. The ethical risk requirements associated with activities and practices to help deliver the minimum marketable features can form ethical risk requirements assessment reports. These reports can help ensure compliance with the necessary regulatory bodies and update the ethical risk management plan \cite{9536679}. 

The last layer represents all the micro activities (technical practices) involved in interpreting the overall strategic vision and identified as user stores (user story) or stories  in an agile context. Stories translate to smaller units of representation and interpretation of the features \cite{rautiainen2011towards}. At this level, individual projects or team activities within the organization and the associated ethical risk requirements are identified as operational ethical risk requirements. These risks form ethical risk requirements lists or registers, which can be used to update the ethical risk requirements assessment report that feeds into preceding layers. Synchronization is required throughout the system.

The ethical stack presents a high-level overview of implementing ethical requirements however, it does not emphasize how management can identify ethical requirements. This serves as a limitation to the framework because one of the critical steps in management involves identifying requirements and assessing them accordingly \cite{freeman2001software}.

\subsubsection{Ethical Risks Requirements:} Cheatham \cite{cheatham2019confronting} explains that while the risks from AI can arise for businesses at any time, having effective mechanisms in place can help mitigate these risks instead of dealing with them on a case-by-case basis or from legislative demands. This argument is also supported by \cite{winfield2019machine}, who encourages using ethical requirements as ethical risk assessments to help reduce the impact of exposure. A risk-based approach can contribute to protecting and preserving the core ethical values within AI or AI-enabled systems by highlighting or identifying ethical risks that require the engagement and commitment of all actors within the organization \cite{9536679}. The perceived risks emanating from the dangers and backlash from the failure of AI systems are considered a factor that may influence the value attached to ethical requirements. Baquero \cite{baquero2020derisking} explains that with the accelerated rate of technological development, the traditional way of managing risks may no longer suffice.

Most companies usually have well-defined risk management plans, but risks associated with AI provide a different kind of complexity. For one, AI can pose an unknown risk by introducing new responsibilities that are less defined \cite{baquero2020derisking}. AI systems with well-developed risk management within a company may interact with data from another agency (human or non-human) to create new risk avenues. The absence of ethical requirements structure in the business's risk management can expose it to new vulnerabilities and risks. AI risks can also be difficult to track across the Enterprise. The occurrence of an incident such as bias can lead to its systematization into the entire organization, which can have catastrophic consequences \cite{baquero2020derisking}. As such, organizations may struggle to implement effective risk management designs without proper AI risk structures from ethical requirements. AI risk management is not a siloed affair that rests solely on executives but requires that all business layers participate as stakeholders.

\subsection{Limitations}
One of the limitations of our study pertains to external validity as all the interviewees are from Finland, giving this study a localized perspective. A broader perspective from other parts of the EU could provide a different insight. While this may pose a limitation, the findings agree with mainstream literature on the management-level approach to implementing ethical requirements. The population size of the interviewees represents a population validity limitation. The interviewee population size could benefit from a more significant number, with a larger population size possibly providing a richer insight into the research. The requirements stack requires additional tools to enable relevant stakeholders to identify which ethical requirements to tackle. 

Future research will focus on developing and testing tools that can enhance the ethical requirement stack in implementing ethical requirements practices within SE organizations. Future research will also investigate the components of risk and sustainability ethical requirements by building business cases on them to examine how they can improve the visibility of ethical requirements at middle-higher level management. 

\section{Conclusion}
The deployment of AI ethics guidelines to serve as ethical requirements for implementing AI ethics in practice has experienced a gradual take-off at the technical or development levels with slow progress at higher management levels. The study aimed to understand how middle and higher-level SE management considers ethical requirements. The empirical process revealed five PECs. PEC1 indicates that ethical requirements are understood in terms of different components and, in some cases, without recourse to the guidelines for trustworthy AI. PEC2 shows that ethical requirements have technical and legal value but no financial value. PEC3 explains how the value of ethical requirements can be improved by harnessing their economic viability and risk assessments. PEC4 examines ethical requirements' current implementation as legal or regulatory instruments, and PEC5 explores risk and sustainability approaches for implementing ethical requirements using simplified or already existing tools and practices. 

One of the key outcomes of the study is how the value of ethical requirements can be improved by harnessing their viability as societal and environmental well-being and technical robustness and safety requirements in the form of risk and sustainability ethical requirements. The use of simplified management practices in the form of an ethical risk requirement stack to implement ethical requirements practically is also explored.

\begin{acks}

This research is supported by The Business Finland funded project Smart Terminals, SMARTER and Ministry of Education funded project AI-Forum.
\end{acks}

\bibliographystyle{ACM-Reference-Format}
\bibliography{Reference}

\end{document}